# Clustering and Trend Analysis of Global Extreme Droughts from 1900 to 2014

**Ehsan Najafi**[1,2,*], **Reza Khanbilvardi**[1,2]

[1]NOAA Center for Earth System Sciences and Remote Sensing Technologies (NOAA-CREST), The City College of New York, The City University of New York, 10031 New York City, USA

[2] Department of Civil Engineering, The City College of New York, 10031 New York City, USA

**Abstract**

Drought is one of the most devastating environmental disasters. Analyzing historical changes in climate extremes is critical for mitigating its adverse impacts in the future. In the present study, the spatial and temporal characteristics of the global severe droughts using Palmer Drought Intensity Index (PDSI) from 1900 to 2014 are explored. K-means clustering is implemented to partition the extreme negative PDSI values. The global extreme droughts magnitude around the world from 1950 to 1980 were less intense compared to the other decades. In 2012, the largest areas around the world, especially Canada, experienced their most severe historical droughts. Results show that the most recent extreme droughts occurred in some regions such as the North of Canada, central regions of the US, Southwest of Europe and Southeast Asia. We found that after 1980, the spatial extent of the regions that experienced extreme drought have increased substantially.

**Keywords**: *PDSI, Extreme Drought, Clustering, Trend Analysis*

**Introduction**

Drought is an extreme climate event with lower than normal precipitation over a period of time (Dai, 2010 ; Sun and Liu 2014) and indefinite start and termination (Bhuiyan, 2004). Drought is classified into different categories such as meteorological drought, agricultural drought and hydrological drought. Meteorological drought for a region is a time span with precipitation significantly below normal conditions and agricultural drought is a period of time with decreasing soil moisture that results in crop losses (Sun and Liu 2014). Hydrological drought is linked with deficits of precipitation that declines streamflow, groundwater levels and reservoir storage (Mishra and Singh, 2010; Abdi and Yasi, 2015). Drought not only has tremendous negative impacts on water resources, but it causes displacement of people (Black et al., 2013), impacts crops (Najafi et al., 2018a) and triggers severe food shortages (Kogan, 1997), malnutrition and famine (Elagib, 2014). Droughts have had the largest negative impacts over the last century (Bruce, 1994). Droughts form gradually and they cannot be easily identified (Unganai and Kogan, 1998). Almost all of the major agricultural lands are susceptible to droughts (Mekonnen and Hoekstra, 2016). Recently, extreme droughts have affected Asia, Europe, Africa, and North America, to name a few (Dai et al., 1998). Unlike floods, the changes in drought characteristics have not been fully investigated (Mishra and Singh, 2010). Climate extremes have uptrended since the last century in many countries across the globe (Asadieh et al., 2016; Poshtiri and Pal, 2016; Poshtiri et al., 2018; Najafi et al., 2016, Najafi et al., 2017, Najafi et al., 2018a, Najafi et al., 2018c, Najafi et al., 2019a, Najafi et al., 2019b). Although future drought projection is uncertain (Burke and Brown, 2008), many climate scenarios, even the conservative ones, confirm that severity and frequency of climate extremes will increase in the future (Parry, 2007). Extreme droughts are infrequent and variable in a changing climate (Dai et al., 1998). Frequency and intensity of extreme droughts have been studied at both local (Kauffman and Vonck, 2011; Yang et al., 2013) and global scales (Dai, 2010; Dai, 2011; Dai, 2012). Extreme droughts have had an upward trend over the past few decades due to climate change (Dai, 2011). Dai (2012) reported increased risk of drought in the twenty-first century. Seneviratne (2012) showed a high uncertainty in global drought trends over the past several decades. Efforts to minimize the adverse consequences of droughts in the future can be optimized (Afshar and Najafi, 2014; Najafi and Afshar, 2015) if knowledge is gained about the past changes and trends of extreme droughts. Though some investigations have been done at the local scale, few efforts have been made to characterize extreme global droughts. Najafi et al. (2018b) characterized extreme droughts across global contents. In this study, we aim to explore how global extreme drought regions have been changed since the beginning of the last century. K-means clustering is used to group extreme droughts across the globe.

---

[*] Corresponding author. Email: enajafi@ccny.cuny.edu

**Dataset and Methodology**

PDSI is used frequently to assess trends of drought (Dai, 2010; Schrier et al., 2011). PDSI, that was originally developed by Palmer in 1965, is the most important index of meteorological drought used for drought monitoring in the US. The simplicity of the PDSI makes it a very suitable tool in drought studies at large scales (Sheffield et al., 2012). To improve the spatial comparability, Wells et al. (2004) proposed a self-calibrating PDSI (SC-PDSI) by calibrating the PDSI using local conditions. SC-PDSI performed better than the original PDSI during the 20th century in some regions such as North America (Schrier et al., 2006). In this study, SC-PDSI been used to evaluate trends and classify extreme droughts across the globe (Dai et al., 2004). Spatial coverage of data is 58.75 south to 76.25 north and 178.75 west to 178.75 east for global land areas on 2.5-degree latitude by 2.5-degree longitude. The PDSI has been successfully applied to quantify the severity of droughts across different climates (Wells et al., 2004). This allows the PDSI to take into account the effect climate change too (Li et al., 2009). The PDSI ranges from about -10 to +10 and values below -3 represent severe drought (Table 1).

**Table 1.** Palmer classification

| extreme drought | severe drought | moderate drought | mild drought | incipient dry spell | near normal | incipient wet spell | slightly wet | moderately wet | very wet | extremely wet |
|---|---|---|---|---|---|---|---|---|---|---|
| -4.0 or less | -3.0 to -3.99 | -2.0 to -2.99 | -1.0 to -1.99 | -0.5 to -0.99 | 0.49 to -0.49 | 0.5 to 0.99 | 1.0 to 1.99 | 2.0 to 2.99 | 3.0 to 3.99 | 4.0 or more |

Out of 7920 grid cells (144 columns and 55 rows), there are 2755 grids with PDSI values. The prominent regions without PDSI data are Greenland and Antarctica. Figure 1 shows the percentage of non-available monthly PDSI data across from 1850 to 2014. After 1948, the number of grids with PDSI data have not changed. In order to study global extreme droughts, the monthly PDSI data from January 1900 to December 2014 are explored in this study. For this mean, the largest negative and 10 largest negative PDSI values of each grid cell over these 1380 months were chosen. Based on Figure 1, some grids do not have data for the entire time span of study. For these grids, the largest negative and 10 largest negative PDSI values were chosen according to their available time span. Figure 2 is the schematic representation of choosing extreme negative PDSI values. In this figure, monthly PDSI dataset are shown as a three dimensional matrix with 144 rows, 55 columns and 1380 layers (for each month from 1900 to 2014), gray boxes depict non-available data.

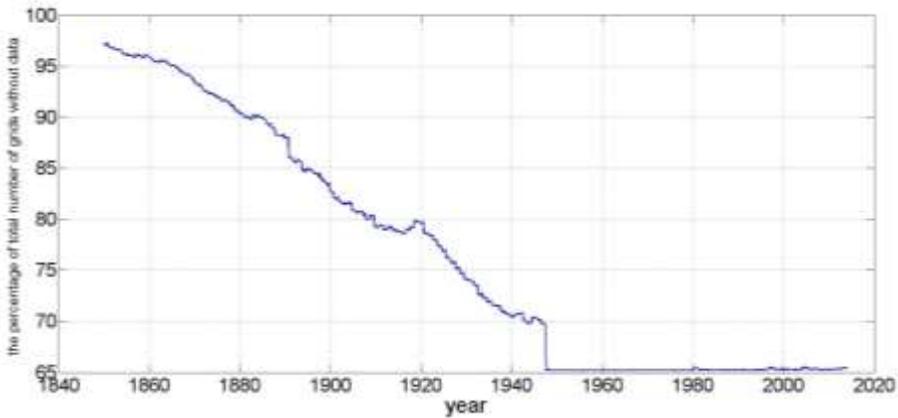

**Figure 1.** The percentage of the total number of grids with no data for monthly PDSI data from 1850 to 2014

K-means clustering method is used to partition the largest negative PDSI values (LNPV). Using this approach, LNPV will be organized into similar groups. K-means clustering aims to partition LNPV into k clusters that share some common trait. In each of these clusters each value is associated with the nearest mean. The K-means clustering algorithm groups randomly chosen centroids. These centroids are beginning points for every cluster. The next step is some repetitive computations for optimizing the centroid positions until centroid locations become stable.

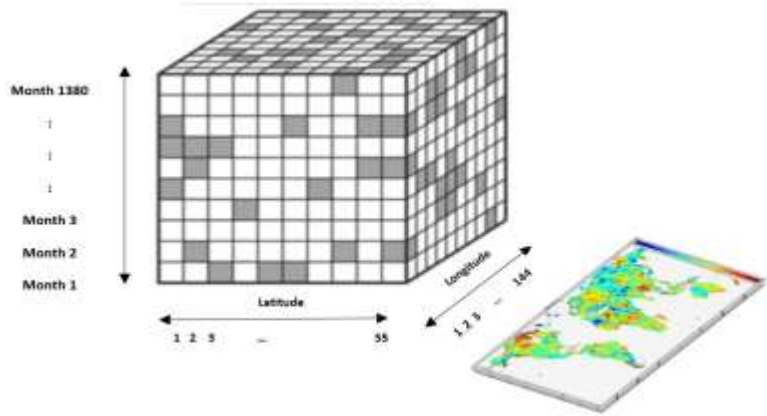

**Figure 2.** Schematic procedure of choosing the 10 largest negative PDSI values

It should be mentioned that like many optimizations techniques, the K-means algorithm can end up with different solutions for different starting point. Silhouette analysis enables us to understand how many clusters are needed.

**Results**

The LNPV and 10 LNPV from 1900 to 2014 are shown in Figure 3. According to these graphs, LNPV are distributed between -8 and -4. All of the grids have experienced extreme drought (PDSI<-4) at least once during the timespan of between 1950 and 1980. LNPV values are generally between -6 and -4, suggesting the extreme drought values were not as severe as the rest of the period. Figure 4 depicts the spatial coverage of the grids that are associated with the 10 LNPV for this time span. These grids are distributed across many regions such as Australia, Europe and Alaska and they do not have any specific pattern. Results show that in 2012, the largest areas around the world experienced their most severe historical droughts. The spatial coverage of these grids are illustrated in Figure 5. Interestingly, many of these grids are located in northern parts of Canada.

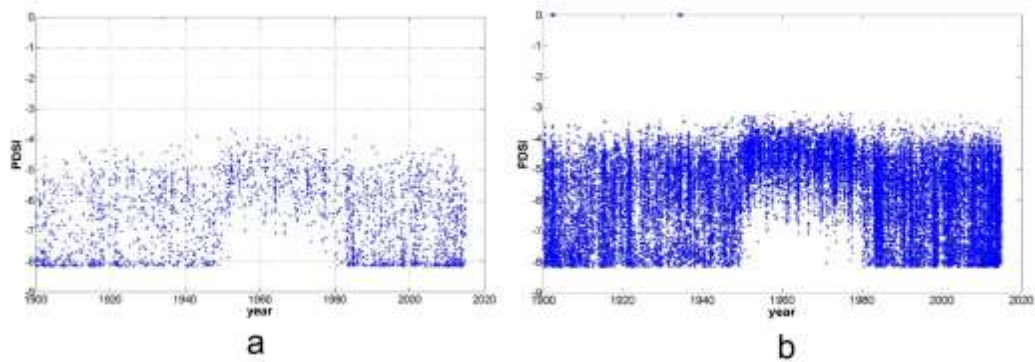

**Figure 3.** (a) LNPV (b) and 10 LNPV from 1900 to 2014

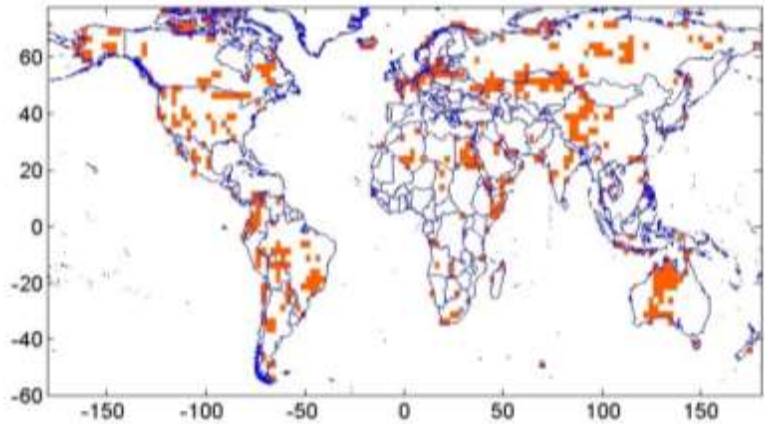

**Figure 4**. Spatial coverage of the grid cells with the LNPV between 1950 to 1980

The most extreme PDSI values and their corresponding dates of occurrence are mapped in Figure 6 and Figure 7, respectively. Droughts occur in virtually all climatic zones (Mishra and Singh, 2010), but severity of the largest negative PDSI in northern latitudes is remarkable. US, many regions of South America, Australia, South Africa, India, and Central Asia represent less extreme PDSI values. In Figure 7 the temporal pattern of these extreme PDSI values shows that extreme droughts in many parts of Canada, most of the Africa, central regions of the US, central Africa, and Southeast Asia occurred during the recent decades. Some countries depict various patterns. For example, extreme droughts in central regions of Russia occurred around 1980, eastern parts of this country experienced its severe historical droughts at the beginning of 20th century though.

Here, we aim to partition the 10 largest negative extreme PDSI droughts from 1900 to 2014. Silhouette analysis is used to choose the number of clusters. The silhouette plot provides a visual method to assess the number of clusters. Figure 8-a demonstrates that 4 is the optimal number for clusters and Figure 8-b displays clusters in 4 different colors that are distributed between 1900, 1930, 1960, 1990 and 2014. The centroids in Figure 8-b are (1978.11, -6.07), (1913.75, -6.87), (1943.53, -6.20) and (2004.34, -6.84).

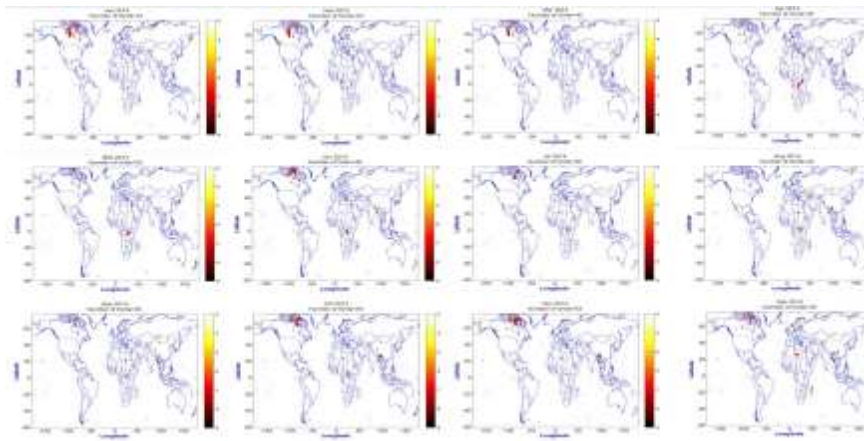

**Figure 5.** Spatial coverage of the grids associated with LNPV in 2012

The boxplots of the 10 LNPV and the associated dates of occurrence are plotted in Figure 9. The distribution of the drought severity in four clusters are very similar to each other. Except date of occurrence, which mainly differ by grids, PDSI discontinuity between 1950 and 1980 is the only reason that make these clusters and the corresponding boxplots different. The spatial coverage of these clusters is mapped in Figure 10. The purple and green grids show the location of the most severe droughts that occurred recently and during the initial decades of 20th century, respectively. The most recent extreme droughts occurred in the North of Canada, central US, Southwest of Europe, Southeast Asia. Conversely, the most severe droughts in some regions of Russia, east of Europe, many parts of Australia, Argentina, etc. have occurred in the beginning of the 20th century.

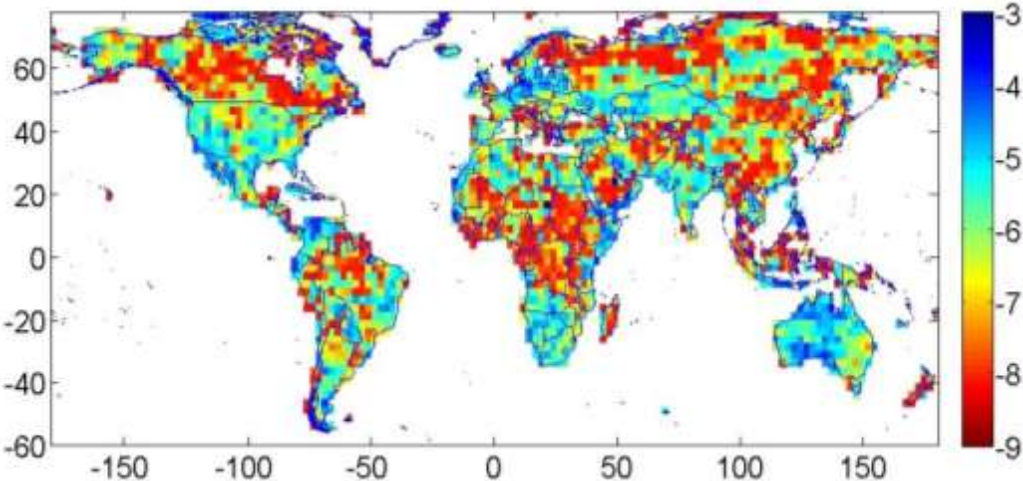

**Figure 6.** Spatial coverage of the LNPV

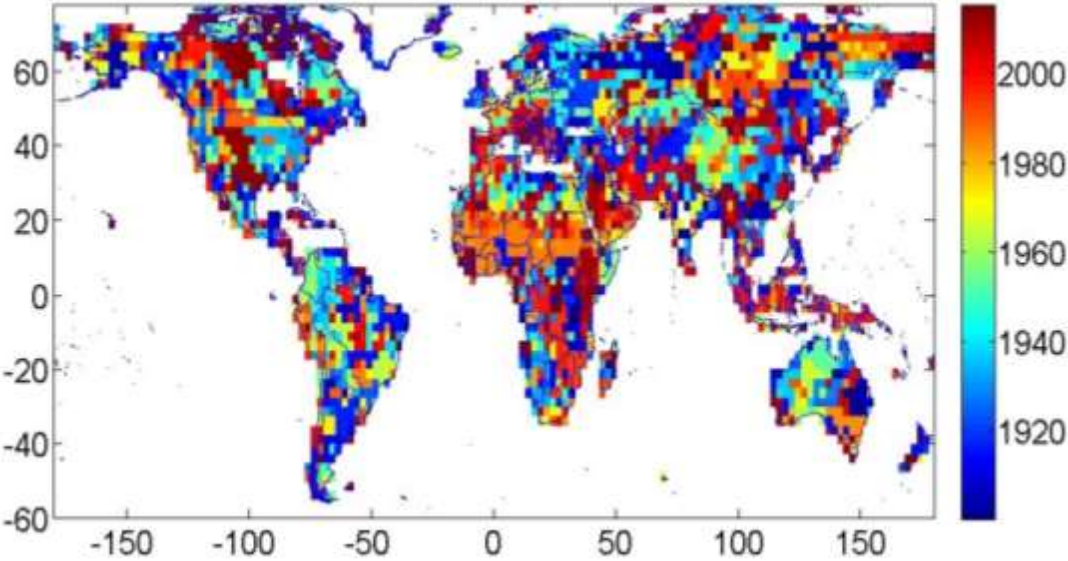

**Figure 7.** Temporal pattern of LNPV

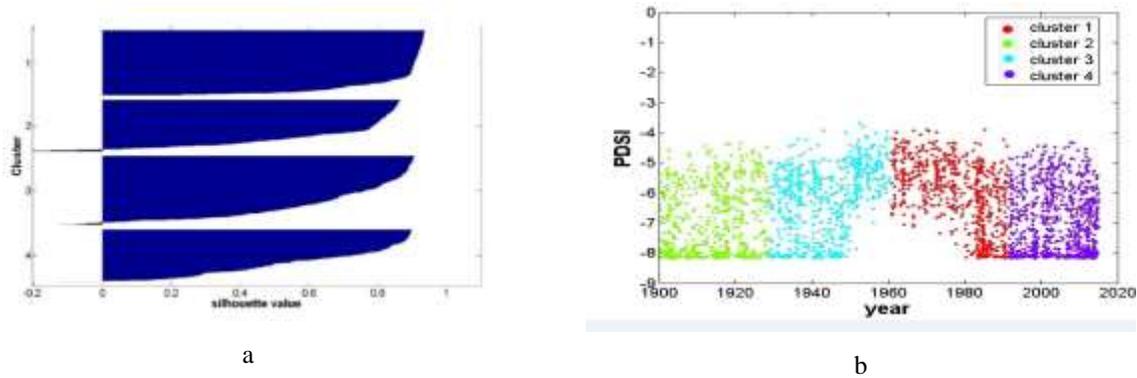

**Figure 8.** (a) Optimal silhouette analysis result with 4 clusters, (b) the LNPV of 2755 grid cells in 4 clusters,

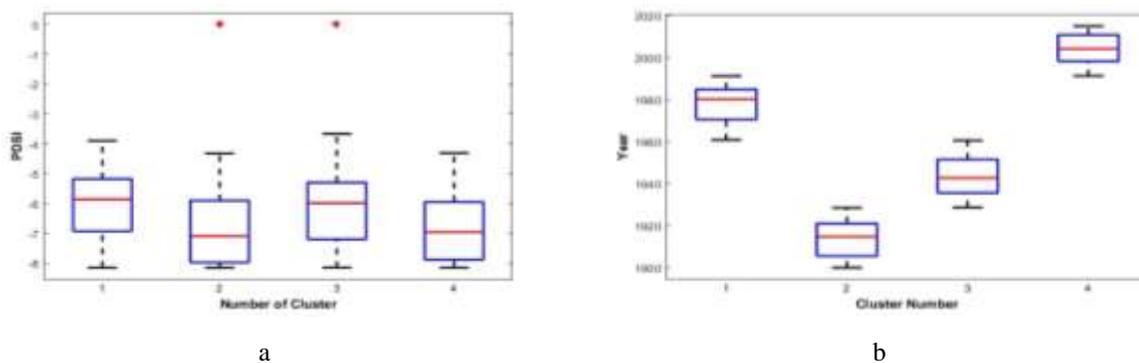

**Figure 9.** (a) Boxplots of 10 LNPV of 4 clusters, (b) boxplots of occurrence time of LNPV of 4 clusters

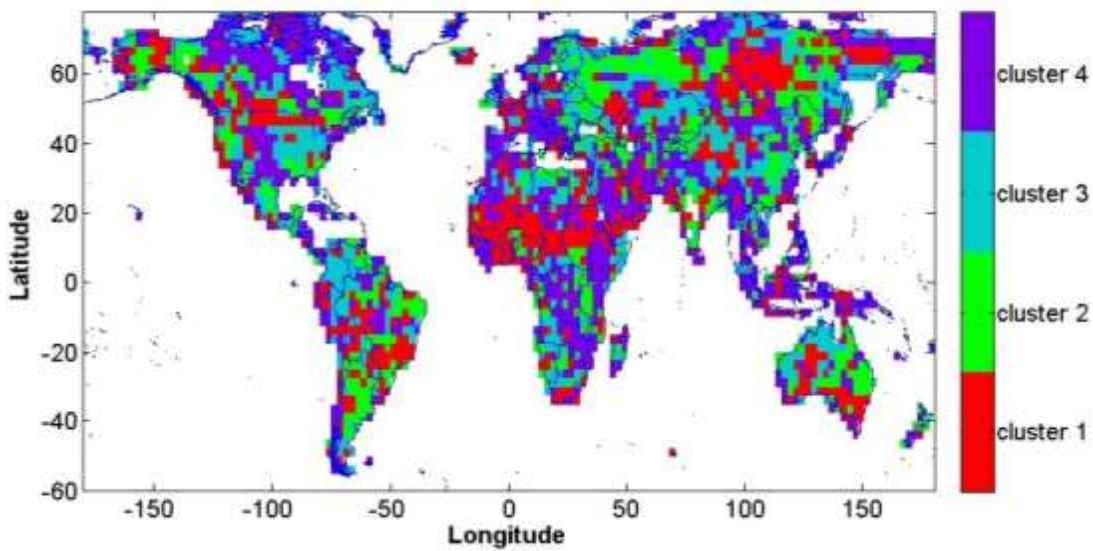

**Figure 10**. Spatial coverage of 4 the clusters

The monthly number of the grids associated with 10 LNPV from 1900 to 2014 are plotted in Figure 11. The linear regression shows an upward trend in this period. The Mann-Kendall test verified this increasing trend. The Mann-Kendall test assesses if a variable has a monotonic upward or downward trend over time. Seneviratne (2012) demonstrated that the total area affected by droughts has increased during the past several years. The annual summation of the grids that are among the 10 LNPV as well as 10, 20 and 30 years moving average are plotted in Figure 12. It is clear that, after 1970, the trend increased steadily. 2012, 1984, and 1983 are the years that many regions around the world experienced their most extreme historical drought (617, 613 and 602 grids, respectively). Of note, it does not necessarily mean that the severity of drought in these years were higher than other years, but, it demonstrates that in these years, more global land areas have experienced their extreme historical droughts. In all the months between 1900 and 2014, except in August 1970, there are some grid cells that experienced one of their 10 largest negative historical droughts. In order to check if the upward trend in Figure 11 is random or not, 95% confidence interval is plotted in Figure 12. As stated earlier, there are 2755 grid cells with PDSI values and the 10 largest negative of their PDSI values were chosen. The probability of occurrence of these 10 numbers in 1380 (1900 to 2014) month is 10/1380=0.0072. For all the 2755 grid cells, 12 random numbers between 0 and 1 are generated and the numbers less than 0.00724 are counted. This process is repeated 100 times and the 5 percentile and 95 percentile of the outcome are chosen. If the whole procedure is repeated 115 (for each year) times and two values of each step (5 and 95 percentiles) of each step are connected together (Equation 1), the outcome will be the two red lines in Figure 12. This procedure is illustrated in details in Figure 13. If the annual trend of 10 LNPV occurs in this interval it can be concluded that by 95 percent confidence, the trend is random.

$$\sum_{i=1}^{i=2755}\sum_{j=1}^{12} N_{ij} \begin{cases} p<0.0072, N=1 \\ p>0.0072, N=0 \end{cases} \qquad \text{Equation 1}$$

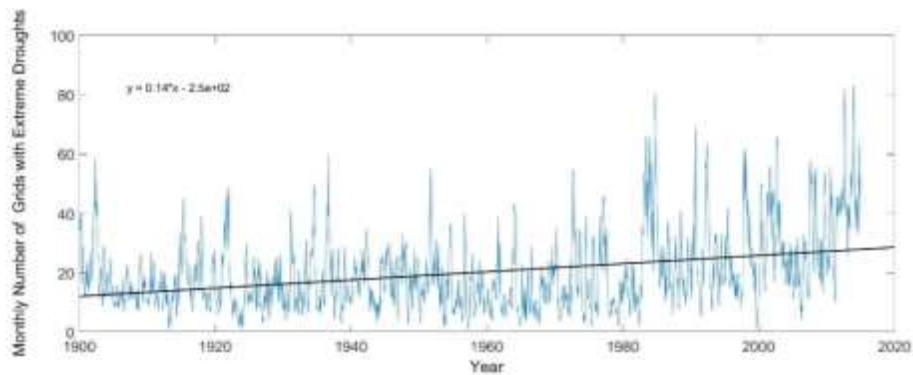

**Figure 11**. The number of monthly 10 LNPV from 1900 TO 2014 and its trend)

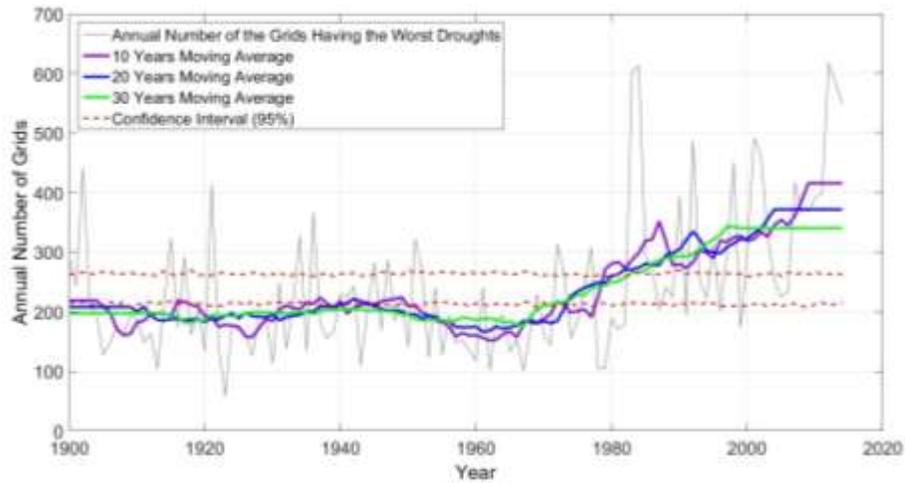

**Figure 12.** 10, 20, 30 years moving average of annual number and 1global land area grids with 10 LNPV from 1900 to 2014

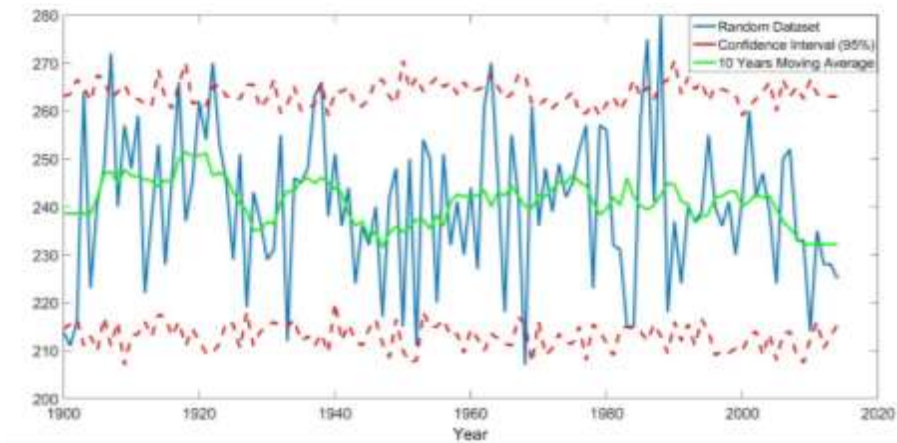

**Figure 13.** Trend of randomly driven extreme drought

Based on Figure 12, from 1900 to 1970 the annual number of grid cells and moving averages, oscillates inside or close to the confidence intervals, but after 1970 the trend crosses the bounds and continues to increase. During the 20[th] century, the average temperature of the planet earth, has occurred in two phases, the first one is from the 1910s to the 1940s (0.35° C), and the second one from the 1970s to the present 0.55° C (Parry, 2007). This growing trend (Figure 12) can be attributed to global warming. Wavelet transformation is a recent development in signal processing that has emerged recently as a tool in trend analysis that explores detailed temporal patterns from both time domains and frequency (Wang et al., 2011). Here, this method is implemented to see whether there is a frequency in annual number of global land area grids with 10 LNPV or not. Results did not show any strong frequency in that time series data.

## Conclusion

Detection, monitoring, and especially prediction of drought is very complicated. In this study, we evaluated the trend of the extreme droughts during the last decades and partition them using K-mean clustering. The results show that from 1950 to 1980, the extreme droughts were less severe across the whole time span. In addition, regions across the globe experienced severe droughts during different episodes. Global drivers of atmospheric variability such as ENSO or NAO influence local climate across the globe (Asadieh et al., 2016; Armal et al., 2016; Armal et al., 2018a, Armal et al., 2018b; Pournasiri Poshtiri and Pal, 2014). The next phase of this study is investigating the association between extreme droughts and large scale climate.

## Acknowledgments

This study was supported by National Oceanic and Atmospheric Administration (NOAA) under Grant CUNY-CREST Cooperative agreement # NA16SEC4810008. The statements contained within the research article are not the opinions of the funding agency or the U.S. government, but reflect the authors opinions.